\documentstyle[aps,prl,psfig]{revtex}

\newcommand{\beq}{\begin{equation}}
\newcommand{\eeq}{\end{equation}}
\newcommand{\beqa}{\begin{eqnarray}}
\newcommand{\eeqa}{\end{eqnarray}}
\begin{document}
\twocolumn[\hsize\textwidth\columnwidth\hsize\csname@twocolumnfalse\endcsname

%\jl{1}

\title{Noisy time series generation by feed-forward networks}

\author{A Priel, I Kanter and D A Kessler}
\address{Department of Physics, Bar Ilan University, 52900 Ramat Gan,
Israel}

%date{\today}

\maketitle

\begin{abstract}
We study the properties of a noisy time series generated by a 
continuous-valued feed-forward network in which the next input 
vector is determined from past output values. 
Numerical simulations of a perceptron-type network exhibit the 
expected broadening of the noise-free attractor, 
without changing the attractor dimension.
We show that the broadening of
the attractor due to the noise scales inversely with the size of the 
system ,$N$, as $1/ \sqrt{N}$.
We show both analytically and numerically that the diffusion constant 
for the phase along the attractor scales inversely with $N$.
Hence, phase coherence holds up to a time that scales linearly with 
the size of the system.
We find that the mean first passage time, $t$, to switch between attractors
depends on $N$, and the reduced distance from bifurcation $\tau$ 
as $t = a {N \over \tau} \exp(b \tau N^{1/2})$, 
where $b$ is a constant which depends on the 
amplitude of the external noise. This result is obtained analytically 
for small $\tau$ and confirmed by numerical simulations.

\end{abstract}
]
%\maketitle

\protect{\vspace*{0.3cm}}

\section{Introduction}\label{chap_int}

The application of neural networks to the field of time series, covers
several areas such as prediction \cite{ts_pred},
identification and control \cite{narendra,ts_anal}. 
The problem of time series prediction was well studied in the past 
\cite{ts_anal} in the context of linear modeling, 
and later was extended to non-linear models.
In this paper we analyze a typical class of architectures used in this field
in the presence of additive noise, i.e.\ a feed-forward network governed
by the following dynamic rule:
\beq\label{dyn_rule}
S_1^{t+1} = S_{out}^t + \mbox{noise} \quad ; \qquad S_j^{t+1}=S_{j-1}^t
\quad j=2, \ldots ,N
\eeq
\noindent
where $S_{out}^t$ is the network's output at time step $t$ and $S_j^t$
are the inputs at that time; $N$ is the size of the delayed input vector.
The focus is set on the long-time (asymptotic) properties of the sequences
generated by the system under the given dynamic rule. The clean model
(without the additive noise) has been investigated \cite{sgen_past,sgen1}
and the main results are summarized below.

Since a realistic time series is noisy, it is imperative to understand the 
effect of noise on the output of the model. In this paper, we conduct an
extensive quantitative study of the effect of noise on this particular
class of model networks. We restrict the analysis to non-chaotic behaviour
for two main reasons. First, chaotic behaviour does not allow long term
prediction due to divergence of nearby trajectories, though such model 
networks are capable of generating chaotic sequences. Second, non-linear
complex (however non-chaotic) time series are an important subclass which
impose interesting questions. Hence understanding the relation between such
complex behaviour and the architecture of the network is crucial form the
point of view of time series prediction.

The basis for using time delayed vectors as inputs is the theory of state
space reconstruction of a dynamic system using delay coordinates 
\cite{takens,embedology}. An architecture incorporating time delays is the
TDNN - time-delay neural network \cite{tdnn}, which when operates in the 
iterative mode contains a recurrent loop (as in the model described above,
without noise). This type of networks is appropriate for learning temporal
sequences, e.g.\ speech signal and for short term prediction. The model we
investigate can be viewed as a degenerate form of a TDNN in which the 
delay-lines are restricted to the input layer. Note that the dynamic rule
(equation \ref{dyn_rule}) corresponds to the closed-loop mode of operation
used for generating subsequent predictions iteratively once the network
has been trained on a given time series. Though some work has been done on
characterization of a dynamic system from its time series using 
neural networks,
not much analytical results that connect architecture and long-time 
prediction are available (see M.\ Mozer in \cite{ts_pred}). Nevertheless,
practical considerations for choosing the architecture were investigated
extensively (see \cite{ts_pred} and references therein).

Recently, it has been shown \cite{sgen1} that an hierarchy among the 
complexity of time series generated by different architectures exists.
This information can be used as a guideline for an application in the 
following way. Given a time series one can conclude some quantitative 
measures regarding the complexity of the sequence, e.g.\ the attractor 
dimension and choose an architecture for the prediction task 
which is high enough in the hierarchy to ensure that it is capable of
generating such a complex sequence.

\begin{figure}
\centerline{\psfig{figure=
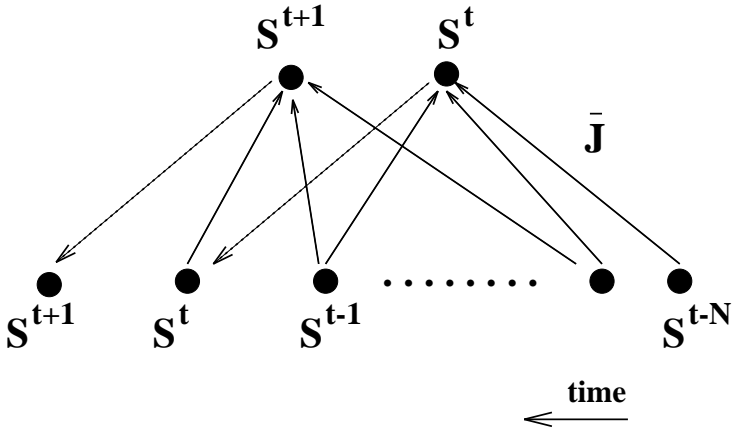,height=6.0cm  }}
\caption{SGen generating a time series. }
\label{sgen_fig}
\end{figure}

Let us review briefly the main findings of the clean model. For 
conciseness we shall refer to the model generating the sequence as a
SGen - Sequence Generator. The simplest SGen consists of a 
perceptron (Figure \ref{sgen_fig}) whose output at time $t$, $S_{out}^t$, 
is determined by the input vector at time $t$,
$~\vec{S}^t~$, $~(\vec{S}^t = (S_1^t,\ldots,S_N^t) )~ $ as follows:

\beq\label{sgen_output}
S_{out}^t=\tanh(\beta {\vec J}\cdot{\vec S}^t)
\eeq

\noindent
for a fixed weight vector $~\vec{J}~$ and gain $\beta$.
The input vector $~\vec{S}^t~$ is given by:

\beq\label{sgen_map}
S_i^t = S_{out}^{t-i} ~,~~ i=1 \ldots N
\eeq

\noindent
i.e.\ the inputs are chosen to be the output values at the previous
$N$ times. Thus, starting from an initial state 
$~S_i=S_i^0 ,~i=1 \ldots N$ 
the system generates the sequence $~S^t ~,~t=1,2,\ldots ~~ $ as
follows:

\beq\label{slp_sgen} 
S^t=\tanh(\beta\sum_{i=1}^{N} J_i S^{t-i})  
\eeq 

%
% ------------------------------------------
% Place Figure 1 around here
% ------------------------------------------
%

In the case of a generic perceptron-SGen, the system is attracted into a 
quasi-periodic (QP) flow governed by one of the Fourier components of
the power spectrum (PS) of the weight vector. 
Hence, the attractor dimension ($AD$) is one.
Denoting the frequency and phase of the governing Fourier component 
by $K$ and $\phi$ respectively, the corresponding part in the
weight vector is
$~J_i = R \cos ({2\pi \over N} K i - \pi \phi) ~~$, 
and the dynamic solution in the leading order of $N$ and 
$~1 \ll K \ll N~$ is of the form:

\beq\label{slp_sgen_sol}
S^t=\tanh \left[ A \cos( {2\pi \over N} (K-\phi) t) \right] 
\eeq 

\noindent 
The amplitude $(A)$ of this solution depends on the gain $(\beta)$ and the 
phase $(\phi)$ in the following way:

\beq\label{slp_sgen_sol_amp}
N \beta = \frac{\pi \phi}{R \sin \pi \phi} 
\left[ \sum_{\rho = 1}^{\infty}
{A^{2\rho-2} \over (\rho!)^2} (2^{2\rho}-1) B_{2\rho}\right]^{-1}
\eeq

\noindent
where $B_{\rho}$ are the Bernoulli numbers. Note that $A$ vanishes 
below a critical value ( which depends both on the amplitude of 
the weight vector $(R)$ and its phase $(\phi)$ )
$~\beta_c = {2 \over R N} {\pi \phi \over \sin(\pi \phi)}~$,
indicating that the system undergoes a Hopf bifurcation at
$~\beta_c~$. 

In the more involved case the model consists of a MLN - Multi-Layer Network. 
The solution is a combination of 
perceptron-like SGen solutions. The exact details, however, 
depend on $\beta$ and the specification of the 
weight vectors (for more details see \cite{sgen1}). 
The $AD$ in the generic case is bounded by the number of hidden units 
connected to the input layer. Moreover, in \cite{nips11} it was shown that
the typical relaxation time for such a system from an arbitrary initial
condition is proportional to the size of the delayed input vector. This 
result is of importance for time series prediction by setting a bound on
the horizon of predictions.

The problem of noise in a dynamic system is of
great importance for the behaviour of the system (e.g.\ stability), 
and hence its implications on the time series measured from that 
system. In the classical theory of time series analysis
(linear and non-linear), one is interested in the prediction ability
of a model when trained with noisy data.
Since one intends to use a SGen to reproduce noisy data, it is
important to understand how noise affects the output of a generic
SGen. In particular, it is crucial that the SGen be robust under the
addition of noise, which is non-trivial given the non-linear feedback
dynamics of the SGen. The addition of noise enables us to check the
stability of the previous results, obtained for isolated models. 

As we shall see, the SGen is indeed extremely stable in the
presence of noise. The noise causes the attractor to broaden. 
Even large noise of order the signal does not destroy the attractor.
This gives rise to several quantitative issues. 
In section \ref{chap_properties} we focus on a 
perceptron-SGen with one Fourier component in its weight vector.
First, we analyze the scaling with $N$ 
of the manner in which the attractor is broadened due to the noise. 
This quantity manifests the cooperative aspect of the degrees of 
freedom in the system. We show that the broadening increases with $N$ 
as $1/\sqrt{N}$. Next, we discuss the issue of phase coherence (PC).
Loss of PC is a generic phenomenon for periodic systems perturbed by 
noise. In this section, we analyze the extent to which 
adding noise to the SGen reduces its PC. The analysis is done for two 
types of dynamic rules, namely sequential updating (described above)
and parallel updating (see section \ref{preliminaries}). 
We show that the phase behaves as a biased random walk process,
as typically observed in noisy oscillators, however the diffusion 
coefficient $D$ exhibits a power-law dependency on $N$. 
For the sequential (parallel) rule, 
$D \sim 1/N^2 ~(1/N)$. The importance of this result is that for large 
systems, PC is lost only over times that scale with the size of the
system. This lost of PC also leads to a broadening of the dominant
component in the power spectrum. We observe that this broadening
decreases with $N$, consistent with the decrease of $D$ with $N$
discussed above.

Next, we measure the $AD$ of the broadened attractor.
As mentioned before, we focus on the classification of various
SGen's by the long term sequence they produce, therefore we are
interested in the estimation of this quantity.
In section \ref{chap_AD} we apply standard methods to estimate
the $AD$ of time series generated by the SGen.
With no noise added, we of course recover the analytical results,
e.g.\ $AD=1$ for the perceptron-SGen. The more important question is
how the noise added to the system influences the measured $AD$.
Our treatment parallels that found in the literature of
dynamic systems where the $AD$ was estimated from a measured (noisy)
time series taken from 
chaotic systems or strange attractors \cite{mizrachi,zardecki}.
We measure the $AD$ of a perceptron-SGen, as well as of a Committee 
Machine whose parameters were chosen in such a way that two Fourier 
components have a non-zero coefficient and whose $AD$, therefore,
should equal $2$.
We found that for length scales greater than the typical size of the 
noise and well below the 
attractor's radius, the $AD$ of the SGen does not differ from the 
expected analytical results. 

Finally, in section \ref{escape_att}
we analyze the effect of noise on a SGen with multiple attractors.
While in the non-noisy case, the perceptron-SGen exhibits a single 
stable attractor, here we expect transitions between attractors due 
to the noise. We focus on the average time needed to escape from a basin
of attraction, and particularly its dependence on the sizes of both the 
system and the attractor. 
This quantity, also known as the mean first passage time,
has been investigated extensively in the context of chemical
reactions, dynamical systems etc. \cite{moss,talkner_1,graham}.
Obviously, we are interested in the case of a discrete system.
This issue has been less treated (see \cite{moss} and \cite{knessl}). 
We consider the case of a system governed by two Fourier components that 
results in two attractors. The problem of escape time is related to 
the evolution of the amplitude in coupled map equations.
The phase portrait of such a map suggests that
the motion in this phase space can be approximated by a one 
dimensional flow of the form:
\beq\label{simple_map}
x_{n+1} = f(x_n) + \xi_n 
\eeq 
\noindent
where $f(x)$ is a non-linear map and $\xi$ is the noise term.
Following the treatment of Talkner et al.\ \cite{talkner_1}, we relate 
our system to the problem of a discrete dynamics with small
non-linearity in the presence of a weak noise. The analytical result
is in a good agreement with extensive simulations of the
perceptron-SGen for both the polynomial prefactor and the leading
exponential part. 

The results presented herein will primarily focus on the perceptron-SGen.
Nevertheless we expect that the general properties and trends remain true
in the more general case.

Summary and a discussion are presented in section \ref{discussion}.

\section{Preliminaries}\label{preliminaries} 

Let us introduce a few concepts which are of general use in the 
following.
The basic model is the SGen in its simplest form -
a perceptron whose output is connected to the first input, as described
in the previous section. This is the {\it sequential} updating rule,
given by eqs. \ref{sgen_output} - \ref{slp_sgen}.

The sequential scheme can be thought of as a fully connected network with
$N+1$ units. The units are updated one at a time, i.e.\ at each time 
step, another unit plays the role of an output unit.
The weight matrix connecting the units is asymmetric with 
a certain spatial structure where the interactions are only a function
of the difference between the location of each pair of units ($ ij $):

\beq\label{toeplitz_matrix}
J_{ij} = W_{i-j~ \mbox{mod}~ N+1}
\eeq

\noindent
where $W_0 = 0$, and $W_l ~ (l \neq 0)~$ is the same weight vector of 
the sequential rule. The main diagonal elements are zero, and the rest
are
the same values as the first row but cyclically permuted, e.g.\ for $N=3$:

\beq
{\bf J}  = \left[	
\begin{array}{cccc}
   0  &   W_1 & W_2   & W_3  \nonumber \\
  W_3 &    0  & W_1   & W_2  \nonumber \\
W_{2} & W_{3} & 0     & W_{1}  \nonumber \\
W_{1} & W_{2} & W_{3} & 0 
\end{array} 
\right]
\eeq

\noindent
This type of weight matrix is said to have a Toeplitz structure.
To implement the {\it parallel} scheme,
all the units are updated simultaneously with the sequential 
rule via the matrix described in equation \ref{toeplitz_matrix}: 

\beq\label{parallel_rule}
S_i^{t+1}=\tanh(\beta \sum_{j=1}^{N+1} J_{ij} S_j^t)  
\eeq

In the sequential scheme, noise is presented to the system in the
following way:

\beq\label{perc_noise_dynamics}
S_1^{t+1} = S_{out}^{t} + \eta^{t} . 
\eeq
where $\eta$ is distributed according to:
\beqa
E[\eta^{t}] & = & 0 . \nonumber   \\
E[\eta^{t} \eta^{t^{\prime}}] & = & \sigma^2 \delta_{t t^{\prime}} .   
\eeqa 

\noindent
In this way, noise is added only to the first unit in each iteration
of the dynamic rule.
In the {\it parallel} updating scheme, the noise is represented by a
vector with $N+1$ independent components ${\vec \eta}^t$,
which is added to all units simultaneously in each iteration: 

\beq\label{noisy_parallel_rule}
S_i^{t+1}=\tanh(\beta \sum_{j=1}^{N+1} J_{ij} S_j^t) + 
\eta^{t}_i  
\eeq

As we previously noted, the sequential SGen produces a time series
which can be denoted by $S^{t}$, $~~ t=1,2,\ldots ~$. 
The sequence $S^t$ is the basis of the numerical analysis. In order to
use the rich theory of reconstructing state space
\cite{takens,embedology}, one has to embed the time series in a
phase space.
The process of embedding a time series onto a {\bf d}-dimensional space,
generates a set of vectors (or a trajectory) in that space.
The embedded vectors are: 

\beq\label{def_embed}
{\bf \vec{X}_{t}} = (S^{t}, S^{t-1}, \ldots , S^{t-d+1}) 
\eeq

\section{Properties of a single attractor}\label{chap_properties}

\begin{figure}
\centerline{\psfig{figure=
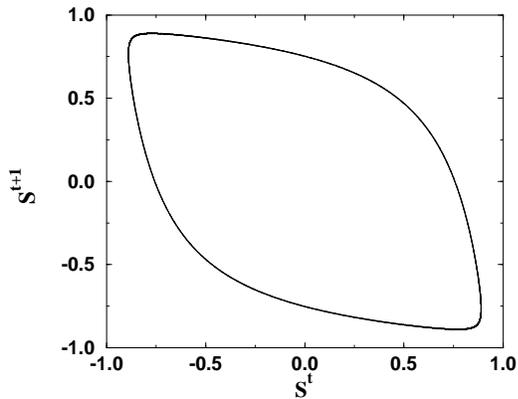,height=6.0cm }}
\caption{Quasi-periodic orbit generated by a perceptron 
$N=50,~ K=17,~ \beta=1/17,~ \phi=.123$.}
\label{qp_orbit_1}
\end{figure}

In this section, we analyze the properties of a perceptron-like SGen
with a weight vector that contains a single Fourier component with an
arbitrary phase $(\phi)$ of the form 
$~~J_j = R \cos \left({2\pi \over N} K j - \pi\phi\right)~$.
When no noise is added to the dynamic equation (equation \ref{slp_sgen}), 
the generic stable solution was found to be a quasi-periodic orbit 
\cite{sgen1}, e.g.\ Figure \ref{qp_orbit_1}. 
%
% ------------------------------------------
% Place Figure 2 around here
% ------------------------------------------
%

When noise is added (equation \ref{perc_noise_dynamics}), the orbit is
broadened. Nevertheless, the system does not become ergodic and the
trajectory is confined in phase space. A characteristic quantity is
the noise induced width of the broadened attractor.
In the following we present both quantitative explanations and
measurements of the dependence of this quantity on the size of the
system $~N$. Next, we discuss the important issue of phase coherence.
A periodic system in the presence of noise typically 
exhibits a loss of phase coherence. This is a result of the fundamental
invariance of the system w.r.t.\ time translation, so there is no 
restoring force to a perturbation which induces a phase shift.
As we shall see, this results in a broadening of the PS of the time
series generated by the system.

\subsection{Attractor broadening}

\begin{figure}
\centerline{\psfig{figure=
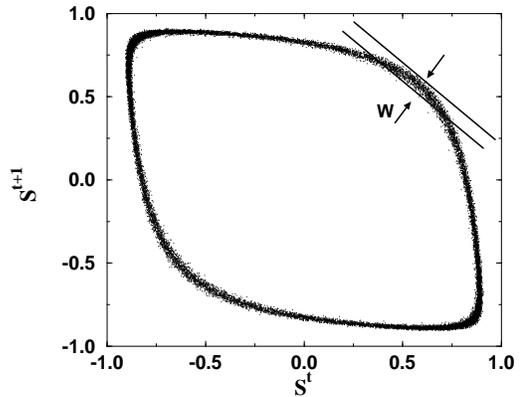,height=6.0cm }}
\caption{Same parameters as previous figure but with a uniform noise of 
amplitude $\pm 0.1$ added.}
\label{qp_orbit_2}
\end{figure}

Let us define the width of the attractor $<W>$ to be the average
local broadening of the embedded time series, see Figure \ref{qp_orbit_2}.
%
% ------------------------------------------
% Place Figure 3 around here
% ------------------------------------------
%

\begin{figure}
\centerline{\psfig{figure=
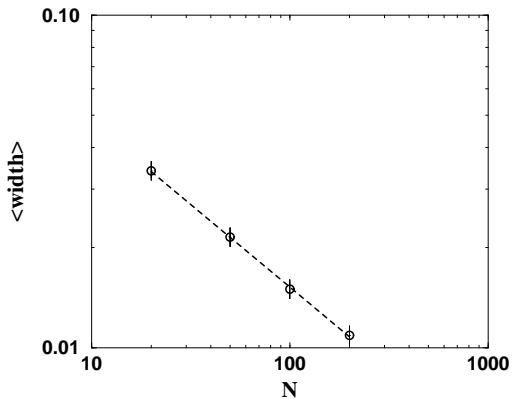,height=6.0cm }}
\caption{The average width of the embedded time series.
The power-law fit (dashed line) is $0.15/N^{\alpha}$ , 
$\alpha=0.5 \pm 0.007$ }
\label{qp_noise_width}
\end{figure}

In this case, we embed the data in a two dimensional space and measure 
the extent perpendicular to the local tangent.
Having done this for a system of sizes $N=20,50,100,200$, 
we plot $<W>$ (denoted by $<$ width $>$ in the figure) vs. $N$ in 
Figure \ref{qp_noise_width}. There exists a clear power-law scaling 
between the two quantities of the form $W\propto A/ \sqrt{N}$ 
where $A$ is a constant. 

% ------------------------------------------
% Place Figure 4 around here
% ------------------------------------------

To understand this scaling law, consider a random vector (RV) in a 
$N$-dimensional space. The relevant quantity is the projection of such 
a RV on a fixed vector - the weight vector $\vec{J}$. 
Denote the output field $h$ as a sum of projections resulting from
the stable solution vector $\vec{S}$ and the noise vector $\vec{\eta}$ :

\beq\label{field_w_noise}
h=\vec{J}\cdot(\vec{S} + \vec{\eta}) \equiv x_s + x_n.
\eeq

\noindent
The components of $\vec{\eta}$ are the last $N$ 
noise terms given by equation \ref{perc_noise_dynamics}.
The output value is then $~S_{out}=\tanh\left[\beta h\right]~$.
In writing equation \ref{field_w_noise} we neglect contributions 
from noise terms after iterations of the map, as these corrections
are proportional to $\beta$ and so are ${\cal O}(1/N)$.
This can be justified as long as the parameter $\beta$ can be written as

\beq\label{beta}
\beta = (1+b) \beta_c
\eeq

\noindent
and $b$ does not scale with $N$.
The term $x_s$ is of $~{\cal O}(N)~$ as this is the exact solution 
without noise. The term $x_n~$ is the focus of our interest.
Since the $\eta_i$ components are RV's ,
we can calculate the first two moments of $x_n$:     

\beqa\label{statistic_of_noise}
E(x_n) & = & \sum_{i=1}^{N} J_i E(\eta_i) = 0 \\
E(x_n^2) & = & \sum_{i,j} J_i J_j E(\eta_i \eta_j) = 
\sum_{i,j} J_i J_j \sigma^2 \delta_{ij} = \sigma^2 \frac{N}{2} \nonumber
\eeqa

\noindent
Thus the variance of the noise term is of ${\cal O}(N)$. The geometrical 
interpretation is that a RV has a projection
which is of ${\cal O}(\sqrt{N})$ on a given direction. 
Since the parameter $\beta$ scales as $~1/N~$ 
(as long as $b$ in equation \ref{beta} does not scale with $N$),
we can conclude that the contribution of the noise
term scales as $~1/\sqrt{N}$, in agreement with
the numerical results presented above. Note that this results hold 
even for large noise values and are linear with $(\sigma)$.

\subsection{Phase coherency}

On general grounds, we expect the phase to undergo a biased random walk,
where the bias represents the frequency of the unperturbed system.
We can measure this directly by comparing the phase of the noisy system
to that of a noise-free ``reference'' system.
Starting from identical initial conditions, the accumulated phase in 
each series is measured.
Denoting the accumulated phase in the clean/noisy series
(subscript $c,n$) at time $t$ by:  

\beqa\label{acc_phase}
\Phi_{c}(t) & = & 
\sum_{i=0}^t \left( \phi_{c}(i+1) - \phi_{c}(i) \right) ~~~~~~
\phi_c(0)=\phi_n(0) \\
\Phi_{n}(t) & = & 
\sum_{i=0}^t \left( \phi_{n}(i+1) - \phi_{n}(i) \right) \nonumber
\eeqa
%
%
% ------------------------------------------
% Place Figure 5 around here
% ------------------------------------------
%
\begin{figure}
\centerline{\psfig{figure=
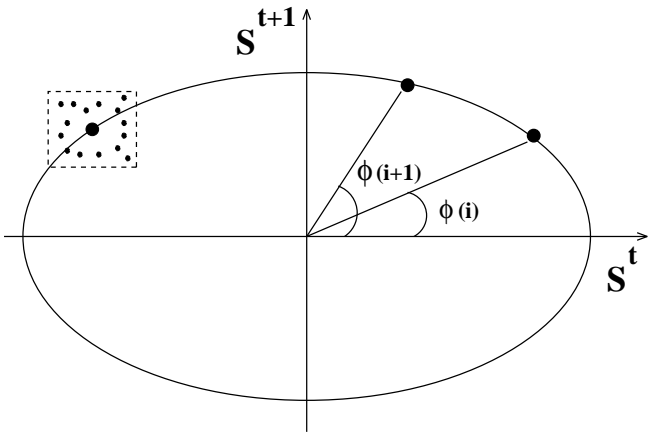,height=6.0cm }}
\caption{Relative phases of the embedded vectors. The left part of the 
figure describes a ``clean'' point surrounded by typical noisy points.}
\label{phase_calc}
\end{figure}

\noindent
where the phases $~\phi_{c}(i)~,\phi_{n}(i)~$ are the relative phases 
of the i'th 
clean/noisy embedded vectors w.r.t. an arbitrary, but fixed, 
coordinate system (see Figure \ref{phase_calc}, ignore the left part of 
the figure).
The quantity of interest is the expectation value of the 
squared phase difference defined by:

\beq\label{phase_diff}
\left<\Delta\Phi^2(t)\right> = E\left[ (\Phi_c(t) - \Phi_n(t))^2 \right]
\eeq

\noindent
where $~<\bullet>~$ stands for the average over all samples taken
after the same time $t$.

\begin{figure}
\centerline{\psfig{figure=
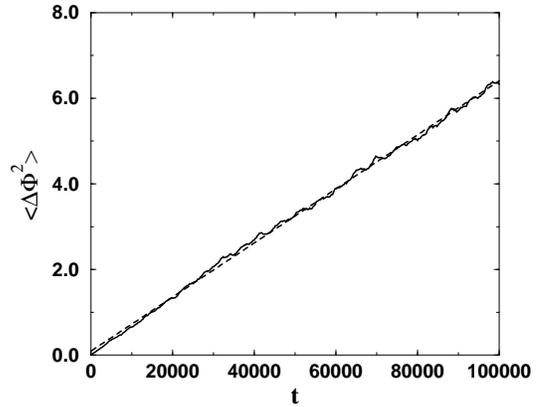,height=6.0cm }}
\caption{Example of the behaviour of the variance of the phase difference 
over time. The slope (dashed line) is the linear regression.}
\label{phase_example}
\end{figure}

An example of the quantity defined in equation \ref{phase_diff} is given
in Figure \ref{phase_example}. 
Clearly, this behaviour indicates that the process is diffusive.
%
% ------------------------------------------
% Place Figure 6 around here
% ------------------------------------------
%

\begin{figure}
\centerline{\psfig{figure=
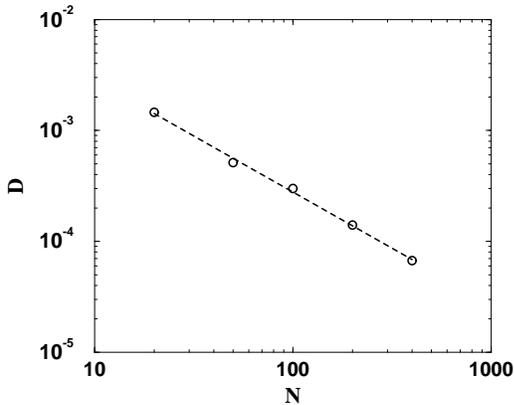,height=6.0cm }}
\caption{Diffusion coefficient for the parallel rule. The linear
regression (dashed line) is 
$~D=0.029/N^a \, \mbox{\,with\,} a=1 \pm 0.03 $.}
\label{phase_par}
\end{figure}

\begin{figure}
\centerline{\psfig{figure=
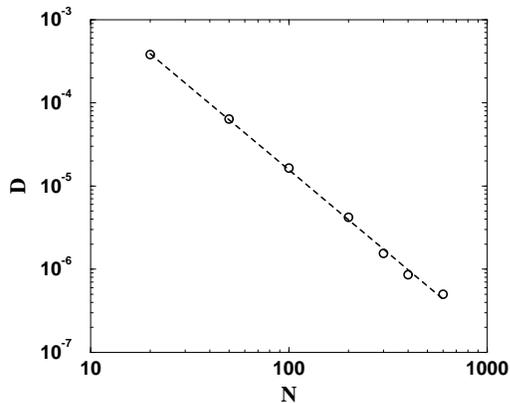,height=6.0cm }}
\caption{Diffusion coefficient for the sequential rule. The linear
regression (dashed line) is
$~D=0.154/N^a \, \mbox{\,with\,} a=2 \pm 0.036 $.}
\label{phase_seq}
\end{figure}

The slope of this figure represents the diffusion coefficient. 
The diffusion coefficient was extracted from data of the type
represented in Figure \ref{phase_example} for both parallel and
sequential updating rules.
Each data point is an average over $400$ samples (as in the figure).
In each case, the simulations were taken at different system sizes. 
The exact 
parameters of each SGen are not important, however they were chosen such 
that the solution is QP and well above the critical value $\beta_c$
where a bifurcation occurs.
Each point in Figures \ref{phase_par},\ref{phase_seq}, is the slope of the
linear regression and the statistical error is less than the size of
the point. The results from
the figures reveal a scaling law of the diffusion coefficient $D$: 
\beq\label{diffusion_coef}
D \sim 1/N^a 
\eeq
\noindent
where $a=1(2)$ for the parallel (sequential) rule.
%
% ------------------------------------------
% Place Figure 7,8 around here
% ------------------------------------------
%

To understand these results,
let us now extend the arguments that led to the ``width'' of the noise
in the previous section. We start with the parallel dynamics and develop
a relation between $~\left<\Delta\Phi^2(t)\right>~$ and time. 
It was shown that the contribution of the noise is of the order 
$~1/\sqrt{N}~$. Examine Figure \ref{phase_calc} (its left part) in the 
context of Figures \ref{qp_orbit_1},\ref{qp_orbit_2}. Each point along the 
clean orbit, is 
surrounded by a cloud whose typical radius is of ${\cal O} (1/\sqrt{N})$.
So basically, the distance between one iteration of the same point in
the clean and the noisy series, is of ${\cal O} (1/\sqrt{N})$.
Since the noise is assumed to be small, the phase can be approximated
by the distance projected on the QP orbit.
Hence, the variance of that phase scales as $1/N$. This explains the
result for the scaling law in the parallel case. 
The sequential dynamics has the same characteristics, however the time
steps should be rescaled w.r.t. the parallel dynamics by a factor of
$1/N$. That is the reason for the $1/N^2$ scaling. \\

One can conclude that phase diffusion indeed occurs (as expected), 
however its associated time scale increases with the size of the system
in a power-law fashion (equation \ref{diffusion_coef}). 
Therefore the system remains coherent over increasingly long times as 
$N$ increases. \\

\begin{figure}
\centerline{\psfig{figure=
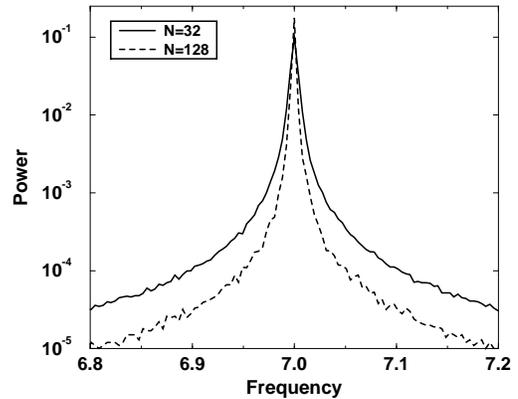,height=6.0cm }}
\caption{Broadening of the dominant component in the power spectrum. The
weight vector consists of one Fourier component with $K=7$. The systems
sizes are $N=32,128$.}
\label{broaden_peak}
\end{figure}

The loss of PC is also manifested in the Fourier domain in the broadening
of the dominant Fourier component. In the unperturbed system, 
the power spectrum of the 
stable solution/state is characterized by a sharp peak (delta function).
The noisy system produces a sequence whose power spectrum
is broadened around the unperturbed Fourier component. The larger the 
phase diffusion constant $D$, the more broadened the dominant component. 
We indeed observe that the broadening decreases with $N$. Figure 
\ref{broaden_peak} depicts the power spectrum of two sequences 
(of the same length) generated by two perceptron-SGen of 
sizes $N=32,128$. The wave number of the single Fourier component 
is $K=7$ and the weight vector is produced according to:
$~J_i = \cos ({2\pi \over N} K i) ~,~~i=1 \ldots N~$.
The power axis is drawn in a log scale to emphasize the broadening
effect.

% ------------------------------------------
% Place Figure 9 around here
% ------------------------------------------

\section{Attractor dimension}\label{chap_AD} 

We have seen that in the case of the perceptron-SGen, the noise
gave rise to a broadening of the attractor. The attractor, nonetheless
remained essentially 1-dimensional, as a perusal of figures 
\ref{qp_orbit_1},\ref{qp_orbit_2} immediately verifies. This is
consistent with the general behaviour of simple attractors in the 
presence of noise. For the case of the MLN, we expect the general
picture to persist. It is however non-trivial to verify this since the
attractor is higher dimensional. We employ for this purpose the tools
that have been developed for analysing dynamical systems 
from their time series. Of course, the question of attractor dimension
is crucial for exploiting these networks for prediction and modeling.

Many methods were proposed for estimating the $AD$. We just mention 
the simplest method, which is the "Box-Counting" \cite{schuster}.
In fact, most methods are based on statistical estimators for the
dimensionality of the attractor. We used the Correlation-Integral method,
that was introduced by Grassberger and Procaccia \cite{grassberger} 
(see also \cite{pawelzik,abarbanel}). In this method the $AD$, denoted
by $D$, is estimated by calculating the correlation sum $C(r)$ from
the data as follows:

\beq\label{correlation_sum}
D=\lim_{\stackrel{N_p \rightarrow \infty}{r \rightarrow 0}} 
\frac{\ln C(r)}{\ln r} 
\eeq

where :

\beq\label{c_r}
C(r)= \frac{1}{N_p^2} \sum_{i,j=1}^{N_p}
\Theta\left(r-\left|\vec{X_i}-\vec{X_j} \right|\right)
\eeq

\noindent
$\vec{X_i}$ are the embedded time series vectors
(equation \ref{def_embed}), $N_p$ is the number of data 
points and $\Theta$ is the Heaviside step function.

In practice, the $AD$ is estimated in the so-called scaling region of
the correlation integral, i.e.\, one has to identify a sufficiently 
large range of lengths scales over which the slope is constant. 
In many cases, the picture is not 
very clear especially when the number of points is not large enough, or
when certain parameters in the algorithm for estimating $C(r)$ are not 
optimized (e.g.\ delay time) \cite{smith,theiler_1}. 
We also note that since the data has a high degree of correlation,
one has to introduce a cut-off to exclude points that were generated
closer (in time) than this value \cite{theiler_1}. This points have
strong correlation that affect the correlation dimension which measures
the correlation between points from different passes of the trajectory.
We used the first zero of the autocorrelation function as a cut-off.

When the measurements are corrupted with noise one can distinguish
between two regimes of length scales; one dominated by the attractor
and the other by the noise. This problem was originally investigated
by Mizrachi et al.\ \cite{mizrachi}, and Zardecki \cite{zardecki}. 
In the broad sense, one can identify four regions \cite{brandstater}.
Due to the finite number of points in the data sample, for very small
$r$, the number of points in the sphere of radius $r$ approaches zero,
and hence also the slope. At larger $r$, there is a transition to a
region where the noise dominates. If the number
of points is large enough, the slope saturates the embedding dimension.
At yet larger $r$, one enters the scaling region with a constant slope 
estimating the $AD$ (given that the region is large enough). 
Finally, the slope returns to zero as $r$ reaches the attractor's radius.
For clarity only the second and third regions are shown.

\begin{figure}
\centerline{\psfig{figure=
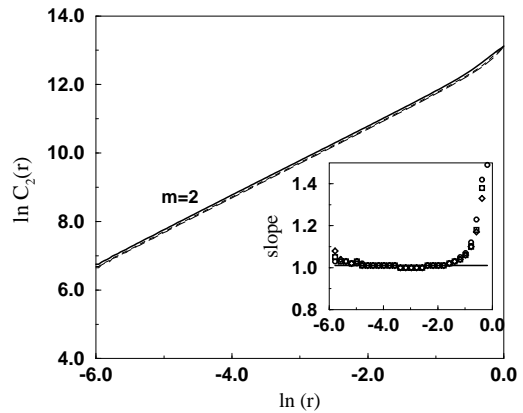,height=6.0cm }}
\caption{Perceptron without noise. $N=400$, $\beta=1/180$, $\phi=0.2235$
$R=1.0$.
The solid guide line in the insert is at $1.01$.
$(\circ ~  m=2,~ \Box ~ m=3,~ \Diamond ~ m=4)$.}
\label{perc_no_noise}
\end{figure}

\begin{figure}
\centerline{\psfig{figure=
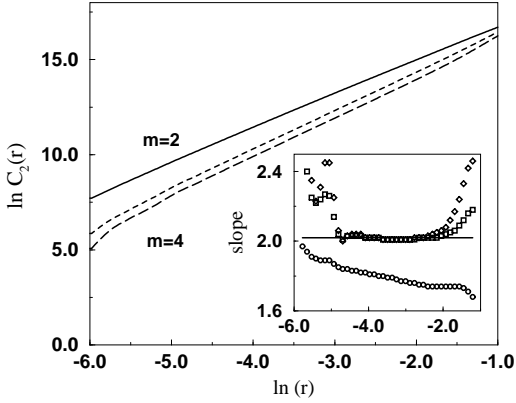,height=6.0cm }}
\caption{CM that exhibits a 2D attractor, without noise. 
$N=500$, $\beta=1/185$, $\phi_i=0.2235, 0.3524, 0.4244.~R_i=1.0~$,
$i=1,2,3$. The solid guide line in the insert is at $2.02$. 
$(\circ ~  m=2,~ \Box ~ m=3,~ \Diamond ~ m=4)$. }
\label{cm_no_noise}
\end{figure}

\begin{figure}
\centerline{\psfig{figure=
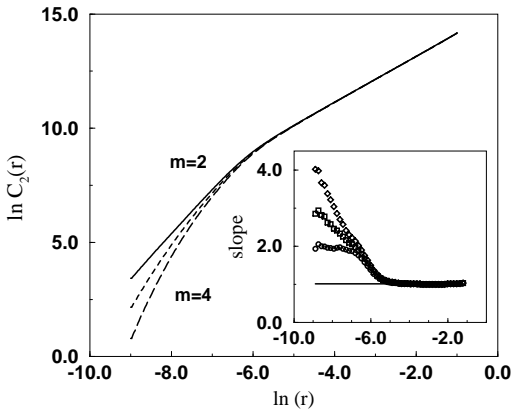,height=6.0cm }}
\caption{Perceptron with noise added. $N=100$, $\beta=1/40$,
$\phi=0.2235$, $R=1.0$.
The solid guide line in the insert is at $1.01$.
$(\circ ~  m=2,~ \Box ~ m=3,~ \Diamond ~ m=4)$. }
\label{perc_noise}
\end{figure}

\begin{figure}
\centerline{\psfig{figure=
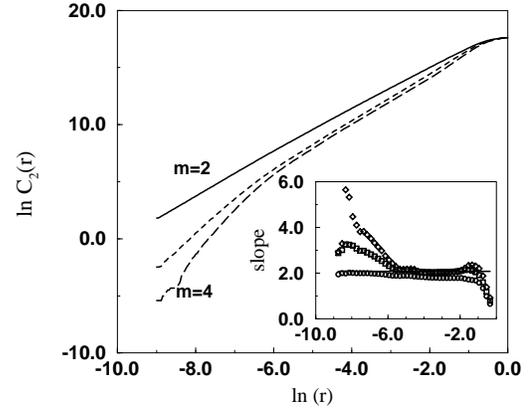,height=6.0cm }}
\caption{CM that exhibits a 2D attractor, with noise added.
$N=500$, $\beta=1/185$, $\phi_i=0.2235, 0.3524, 0.4244.~R_i=1.0~$ , 
$i=1,2,3$. The solid guide line in the insert is at $2.07$. 
$(\circ ~  m=2,~ \Box ~ m=3,~ \Diamond ~ m=4)$. }
\label{cm_noise}
\end{figure}

Let us now describe our measurements.
The time series were generated for four cases:
a perceptron without noise (Figure \ref{perc_no_noise}); perceptron with
noise added as described in equation \ref{perc_noise_dynamics} 
(Figure \ref{perc_noise}); and a Committee Machine (CM) with three hidden
units with and without noise (Figures \ref{cm_no_noise},\ref{cm_noise}). 
By a CM we mean a two-layered network whose
second layer weights equal one. Each perceptron in the hidden layer 
(as well as the perceptron-SGen) has only one Fourier component in 
its weight vector and an arbitrary phase :

\beq\label{weight_1fc}
J_j^h = R_h \cos \left({2\pi \over N} K_h j - \pi\phi_h\right)
\eeq

\noindent
where $R$ is an amplitude, $N$ is the input size, $K$ is the wave number,
$\phi$ is a constant phase shift and $h$ labels the hidden unit. 
( The case of more than one Fourier component is treated in a different
context in section \ref{escape_att} ).
The gain parameter $\beta$ in the
CM was chosen so that the stable attractor of this SGen contains only two
components in the power spectrum. This choice produces a $2D$ attractor.
(The values of all the parameters are given in the figure captions).

The figures present the calculated $~~\ln C_2(r)~$. The $AD$ is 
estimated by the local slope, \, $d[ \ln\,C_2(r)]/d[ \ln\,r]~~$  and
presented in the insert.
It is important to note that all data points are rescaled to the region 
$[0,1]$, prior to the evaluation of the correlation integral.
Figure \ref{perc_no_noise} presents results for the simplest 
perceptron-SGen with only one Fourier component in its weight vector.
The arbitrary chosen phase shift $\phi$ results in a QP
orbit \cite{sgen1} which is $1D$ ($AD=1$). 
We embedded the time series in $m=2,3 ~\mbox{and}~ 4$ dimensional spaces.
Clearly the measurements support the analytical results and the $AD$ 
measured is about $AD=1.01$. 
In Figure \ref{cm_no_noise}, we present the results for the more 
complicated attractor generated by the CM. 
The expected $AD$ is $2~$ (as described above). 
The results are slightly above 2, that is $2.0< AD <2.03$.
Notice that the embedding in a 2D-space gives a wrong result, 
as expected, since the structure of the attractor is unfolded only 
in a 3D space, at least. 

Now we analyze the same perceptron-SGen and CM but with noise added 
(see Figures \ref{perc_noise}, \ref{cm_noise}).
We embedded the time series as in the non-noisy case in $m=2,3$ and $4$
dimensional spaces. We used a uniformly distributed noise with an 
amplitude of $\pm 10^{-2}$ while the attractor's amplitude is bounded
by $\pm 0.7 (\pm 1.2)$ for the perceptron (CM), prior to rescaling. 
Our results are similar to other noisy dynamic systems \cite{mizrachi}
in the sense that for length greater then the characteristic noise
scale, the measured $AD$ saturates the true dimensionality,
i.e.\ in this case $AD=1 (2)$ (as in the non-noisy case). However below
that scale, the noise dominates and since in general it fills the
space in all dimensions, the slope increases with the embedded
dimension. In our case, the slope measured for the noise is correct
only in $m=2,3$, while in higher dimensions it is lower than the
embedded dimension. The reason for this inaccuracy is that we have
not used enough points so the space was not filled densely by the noise. 
The results are $AD\sim 1.01 (2.07)$ which are slightly higher than the 
non-noisy case for the CM.
% ------------------------------------------
% Place Figures 10..13 around here
% ------------------------------------------

In all the figures, one can easily identify the scaling region which is 
quite broad, more than an order of magnitude of length. 
The conclusion from the results is that the SGen maintains its $AD$ in
the presence of noise. The effect of noise is bounded to small length
scales, as expected.

\section{Escape from a meta-stable attractor}\label{escape_att}

So far, we have discussed several properties of the dynamics in 
the neighborhood of a single attractor. 
This section is devoted to the analysis of the dynamics
when there are multiple attractors.
In particular, we focus on the 
average time to escape from the domain of one of the attractors.
The picture one should have in mind is of several states having
local stability with transitions between them induced by noise.

In the first section we derive an analytical result for the mean first
passage time (MFPT) in the limit of weak noise and a weakly non-linear
map. The reasons for taking these limits will be explained later.
The second section describes a series of simulations which support
the analytical results.

\subsection{The mean first passage time in periodic attractors}

In the following we analyze the case where
each of the meta-stable states is characterized by an N-states periodic
attractor. This property is achieved by setting the phase shift $\phi$
(e.g.\ equation \ref{weight_1fc}) to zero \cite{sgen1}. 
In order to keep the discussion
as simple as possible, let us restrict ourselves to the case of a 
perceptron-SGen
with two Fourier components in the power spectrum of the weight vector.
Hence, the weight vector is given by:
\beq\label{weight_two_fc}
J_j = \sum_{m=1}^{2} R_m \cos ({2 \pi \over N} K_m j)
\eeq

\subsubsection{A simplified model}

The key point is our ability to identify a low dimensional discrete 
dynamics that describes the evolution of the solution, 
and relate it to our problem of the SGen. 
In \cite{sgen1} it has been shown that the general solution for a 
perceptron-SGen with a weight vector defined by equation \ref{weight_two_fc},
is of the form:

\beq\label{solution_two_fc}
s^t = \tanh \lbrack \sum_{m=1}^2 A_m  
\cos (~{2 \pi \over N} k_m t ~) \rbrack 
\eeq

\noindent
This solution leads to self-consistent coupled equations for 
the amplitudes of the dynamic solution:

\beq\label{coupled_eq}
A_m^{n+1} =   \beta N R_m  
\sum_{\rho = 1}^{\infty} C(\rho) 
\sum_{t=0}^{\rho-1} 
{(A_m^n)^{2t+1}(A_{m^{\prime}}^n)^{2\rho-2t-2}
\over (t+1)!t!(\rho-t-1)!^2}
\eeq 

\noindent
where $n$ labels the discrete time, 
$C(\rho) \equiv (2^{2\rho}-1) B_{2\rho}/\rho$ 
($B_{\rho}$ are the Bernoulli numbers),
and $m^{\prime}=2$ for $m=1$ and vice-versa.

In the absence of noise, the coupled equations evolve into one of the
two fixed points (f.p.'s) in which only one of the Fourier components
has a non-vanishing coefficient.
The addition of noise, as described in equation \ref{perc_noise_dynamics},
generates a perturbation in each of the coupled equations.
The perturbation can ``kick'' the system out of the vicinity of one
stable f.p.\ so that it escape to the other f.p.
We are interested in the mean time for such an event to occur.

We assume $R_1=R_2=1$, i.e.\ the symmetric case.
In order to continue, we truncate and transform the coupled 
equations (equation \ref{coupled_eq}). 
For small amplitudes, one need keep terms only up to third order.
The result becomes: 

\beq\label{coupled_eq_trunc}
A_m^{n+1}  = \beta \frac{N}{2} \left[ A_m^n - \frac{1}{4} (A_m^n)^3 - 
\frac{1}{2} A_m^n (A_{m^{\prime}}^n)^2 \right]
\eeq

\noindent
where, as before, $m^{\prime} =2$ for $m=1$ and vice-versa.
One can treat these equations as a recursive solution for the amplitudes 
of the dynamic solution.
In this sense, equations \ref{coupled_eq_trunc} become discrete dynamic
equations. For notational convenience we shall relabel the variables
with $A_1 \rightarrow x$ and $A_2 \rightarrow y$.
In addition, we introduce the reduced variable $\tau$ as follows:

\beq\label{tau}
\tau = \frac{\beta - \beta_c}{\beta_c} ~~~ \Longrightarrow ~~~
\beta \frac{N}{2} = 1 + \tau
\eeq

\noindent
where $\beta_c = 2/N $. This redefinition allows us to rewrite 
equation \ref{coupled_eq_trunc} as an $N$-independent map:

\beq\label{coupled_eq_fin}
x_{n+1} = (1+\tau) \left[ x_n - \frac{1}{4} x_n^3 - 
\frac{1}{2} x_n y_n^2 \right] 
\eeq

\noindent
The second equation is obtained
by replacing $x$ by $y$ and vice-versa, $x \Leftrightarrow y$ .

\begin{figure}
\centerline{\psfig{figure=
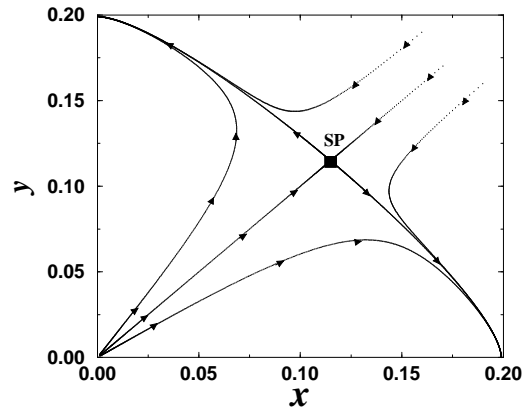,height=6.0cm }}
\caption{Phase portrait of the $2D$ map with $\tau=0.01$. 
SP denotes a saddle point.} 
\label{phase_portrait}
\end{figure}

Analysis of these equations under the assumption that 
$\tau \ll 1$, gives four symmetric f.p.'s, namely: 
$ y^{\star} = 0 ~,~ x^{\star} = \pm 2 \sqrt{\tau}$ and vice-versa. 
These f.p.'s are stable and we consider only the positive ones.
In addition, we have a trivial unstable f.p.\ and four saddle points at
$~ x_{sp}^{\pm}, y_{sp}^{\pm} = \pm {2 \sqrt{3} \over 3} \sqrt{\tau} ~$.
A typical phase portrait of this map is depicted 
in Figure\ \ref{phase_portrait}
(actually, only the positive quadrant is shown). The stable f.p.'s are at
$~x^{\star} = 0.2 ~,~ y^{\star} = 0~$ and 
$~y^{\star} = 0~,~x^{\star} = 0.2~$
( the other two symmetric f.p.'s are not shown ). The saddle point shown 
is at
$~ x_{sp}^{+}, y_{sp}^{+} = {0.2 \sqrt{3} \over 3} ~$, whereas the 
other three are not shown. Let us denote this point by $~SP^{+}~$,
and with $~SP^{-}~$ denoting  the other saddle point, i.e.\ 
$~( x_{sp}^{+}, y_{sp}^{-} )~$.
%
% ------------------------------------------
% Place Figure 14 around here
% ------------------------------------------
%

\noindent
The boundary between the two domains of attraction is clearly the line
$x=y$. 
The additive noise, as mentioned above, perturbs these equations and 
as a result the system may escape from the domain of attraction defined
by $x>y$. The random time it takes for the system to reach the state
$x=y$ is the first passage time stochastic variable. Note that the 
additive noise described in equation \ref{perc_noise_dynamics} is not the 
same one used in 
our model here, since the first noise is applied directly to the SGen, 
while the second is the effect of such noise on the amplitude of the 
solution. The connection between the following model and the SGen is 
given in the next subsection.

The model for the perturbed system, is described by the following 
$2D$ noisy map:

\beq\label{2D_noisy_map}
x_{n+1} = (1+\tau) \left[ x_n - \frac{1}{4} x_n^3  -\frac{1}{2} x_n y_n^2
\right] + \xi_n .
\eeq

\noindent
where $\xi_n$ is a Gaussian additive noise distributed according to:

\beq\label{noise_dist}
\rho_{\epsilon} (\xi) = \frac{1}{(2 \pi \epsilon)^{1/2}} 
\exp - \frac{\xi^2}{2 \epsilon}
\eeq

\noindent
The map for $y$ is obtained in the same manner as in equation
\ref{coupled_eq_fin}.
Note that due to the mutual independence of the $\xi_n$, 
the process defined in equation~\ref{2D_noisy_map} is a Markov process.

The region of interest is of ${\cal O}(\sqrt{\tau})$. Following an 
appropriate rescaling of equation~\ref{2D_noisy_map}, we get:

\beq\label{2D_scaled_map}
\tilde{x}_{n+1} = f(\tilde{x}_n,\tilde{y}_n) + \tilde{\xi}_n  = 
\tilde{x}_n - \tau \left[ - \tilde{x}_n + \frac{1}{4}\tilde{x}_n^3 + 
\frac{1}{2} \tilde{x}_n \tilde{y}_n^2 \right] + \tilde{\xi}_n ~ .
\eeq

\noindent
where $\tilde{x}_n, \tilde{y}_n, \tilde{\xi}_n $ are the rescaled
variables. We further rewrite the map in the following way:

\beq\label{2D_function_map}
f(x_n,y_n) = x_n - \tau {\cal U'} (x_n,y_n).
\eeq

\noindent
The derivative is taken with respect to $x$ or $y$, depends on the
variable for which the map is written for.

Say the initial condition is $y=0 ~,~ x=x^{\star}$, i.e.\ one of the
f.p.'s. Since the line connecting this f.p.\ and the saddle point
is a valley, we may assume that the most probable escape route is
along this line (or its mirror through the x-axis, i.e.\ the line
connecting the f.p.\ with
the saddle point $( x_{sp}^{+}, y_{sp}^{-} )~$ ). This argument can be 
understood by rotating each noise term tangent and perpendicular to the 
path. The perpendicular term decays fast due to the restoring force, 
hence we can conjecture that the dynamics is mainly $1D$. 
Therefore, with the assumption of weak noise and $\tau \ll 1$ we can
reduce the map into one dimension, on that path
(for details, see \cite{next_paper}). Hence, a $1D$ noisy map is
obtained:

\beq\label{1D_noisy_map}
s_{n+1} = s_n - \tau {\cal U'} (s_n) + \hat{\xi}_n ~ .
\eeq

\noindent
where $s$ defines the path. The noise term now is the tangential
projection of the $2D$ noise on the path.
The path $s$ can be found by writing an equation for $x(y)$ on the path,
however the relation is implicit and cannot be used directly.

This type of a $1D$ equation has been investigated for the case of small 
non-linearity \cite{talkner_1,knessl,talkner_2}, namely, the class 
of map functions 
with the property that $f(x)$ deviates only weakly from the identity map:

\beq\label{map_class}
f(x) = x - \tau \frac{dU(x)}{dx} ~,~~~~\tau \ll 1
\eeq

\noindent
The analogy with our $1D$ map (e.q.\ \ref{1D_noisy_map}) is obvious. 
In the next section, we adapt the
derivation of \cite{talkner_2} to our map.

\subsubsection{MFPT analysis}

In the following, we sketch the calculation of the MFPT for the process 
defined in equation~\ref{1D_noisy_map}. The complete derivation and
simulations will be given in \cite{next_paper}.

Assume that the process described in equation~\ref{1D_noisy_map} is defined 
in $(- \infty , \infty )$ and define the random variable $\tilde{t}(s)$,
the first passage time from the interval 
$I= \left[ SP^{-}, SP^{+} \right] $ , by:
\beq\label{fpt}
\tilde{t} = \min \{ n : |s_n| \geq s_{sp}^{+} \} ~.
\eeq
\noindent
i.e.\ the first time the process hit one of the boundaries,
where $SP^{\pm}$ are the saddle points defined above, and 
$s_{sp}^{+}$ is the value of $s$ at the saddle point.
The MFPT, $t(s)$, starting from a point in $I$ is given by:
\beq\label{mfpt_def}
t(s) = < \tilde{t} (s) >  = E[~\tilde{t} ~|~ S_0 = s~] . 
\eeq
\noindent
It was shown that the MFPT can be written as (e.g.\ \cite{talkner_2}):
\beq\label{mfpt_int}
t(s) - 1 = \int_{I} P(z|s) t(z) dz
\eeq
\noindent
where $~P(z|s)~$ denote the transition probability to go from $s_n=s$ to
$s_{n+1}=z$ in a single step.
Under the assumption of weak noise $\epsilon \ll 1 $, the function
$t(s)$ is nearly constant inside the domain of attraction. Fluctuations 
occur mainly near the boundary. The reason is that only close to the 
boundary may one have a finite probability to jump over the 
boundary in small number of steps. Therefore, it was suggested 
\cite{knessl,talkner_2} that this function be written as a product of 
a constant value, and a boundary layer function:
\beq\label{boundary_function}
t(s) = T \tilde{h}(s) ~~~~,~~ \tilde{h}(s^{\star}) = 1
\eeq
\noindent
where $s^{\star}$ is the f.p.
The boundary layer extends a distance of order ${\epsilon}^{1/2}$ around 
$s=s_{sp}^{+}$, and we can write the scaled boundary layer 
function $h(s)$,
$h(s)=\tilde{h}((2\epsilon)^{1/2}s)$. Inserting this assumption in 
equation \ref{mfpt_int} gives an integral equation for $h(s)$.
This equation was analytically solved by Talkner et al.\ 
\cite{talkner_2} and by Knessl et al.\ \cite{knessl}. 
The leading exponential part of the solution of this equation gives:
\beq\label{lead_exp}
T \propto \exp \left[ \frac{2 \tau}{\epsilon} ( {\cal U}(SP^{+})
- {\cal U}(s^{\star}) ) \right ]
\eeq
\noindent
The potential difference has been calculated analytically (see 
\cite{next_paper}) and found to be 
$~{\cal U}(SP^{+}) - {\cal U}(s^{\star}) = {1 \over 3} $.
\noindent
The prefactor is obtained from integrals involving the boundary layer
function. The final result for the MFPT reads:
\beq\label{fin_mfpt}
T = \frac{a}{\tau} \exp \left(\frac{2}{3} \frac{\tau}{\epsilon} \right )
\eeq
\noindent
with $a$ constant. 

\begin{figure}
\centerline{\psfig{figure=
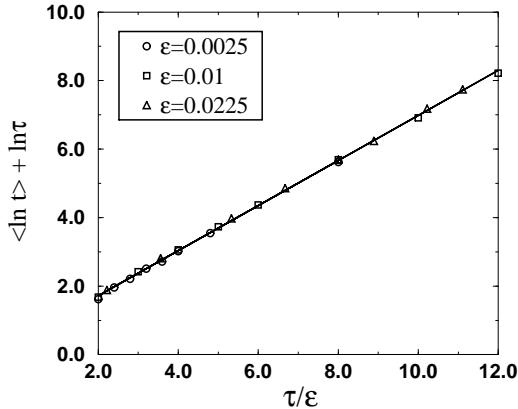,height=6.0cm }}
\caption{Scaling of the average logarithm of the escape time in the
$2D$ model. 
The solid line is a linear 
regression and its slope is $0.658 \pm 0.003$.}
\label{sim_2d_fig}
\end{figure}

Simulations of our $2D$ model (equation \ref{2D_scaled_map}) are shown in
Figure\ \ref{sim_2d_fig}. The reduced variable $\tau$ is varied for
different noise amplitudes $\epsilon$. The results are in excellent
agreement with the prediction of the $1D$ theory (equation \ref{fin_mfpt}).

% ------------------------------------------
% Place Figure 15 around here
% ------------------------------------------

In the next sub-subsection we present the results from
extensive simulations of the real system, i.e.\ the SGen.

\subsection{Numerical simulations}

Measuring the MFPT directly from the time series generated by the noisy
SGen is impossible, since there is no way to distinguish between the 
different attractors. The natural variable which does measure the 
projection of the current state on each attractor is the relative
amplitude in the power spectrum of the input vector. Note that there
exists an equivalence between the amplitudes of the solution to the
coupled equations (equation~\ref{coupled_eq}) and the amplitude in the
power spectrum of the corresponding Fourier components.

We study an SGen with a weight vector containing two Fourier
components, as described at the beginning of the previous section, 
with no phase shift and both amplitudes equal to one, $R_1=R_2=1$. 
We applied the sequential updating scheme described in 
section \ref{preliminaries} with a noise which is normally distributed
${\cal N}(\mu = 0, \sigma=0.1)$.
We set the initial conditions for each run to one of the Fourier
components. In each experiment we measure the number of iterations
before the amplitudes of the two components in the power spectrum of
the input vector, become equal.
As we expect an exponential behaviour of this quantity, we record the 
logarithm of the first passage time. We found that actually the average 
logarithm of the median first passage time has smaller variations 
than the average logarithm of the first passage time over all the data
set. Each pair $(N,\tau)$ was tested $200-400$ times and the first
passage time was recorded. The list of times was divided into $10$
groups and the average logarithm of the median from each group was
taken. Finally, we end up with $10$
values from which we calculated the first and second moments.

\begin{figure}
\centerline{\psfig{figure=
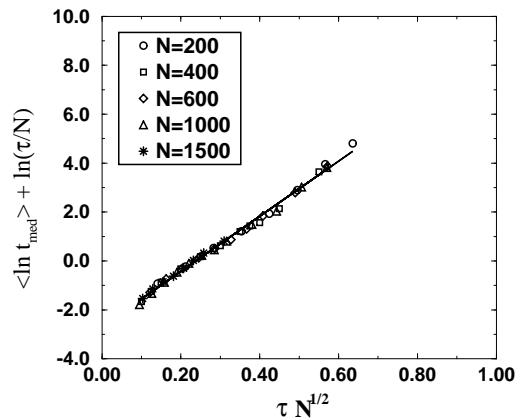,height=6.0cm }}
\caption{Scaling of the average logarithm of the median escape time. 
The solid line is a linear 
regression and its slope is $11.2 \pm 0.12$.}
\label{escape_scale}
\end{figure}

Figure \ref{escape_scale} depicts the ensemble of all experiments, in which
we varied the size of the system in the range $200 \leq N \leq 1500$
and the reduced variable $\tau$ in the range $~ 0.003 < \tau < 0.04~$. 
To demonstrate the scaling
properties, we plot the average logarithm of the median escape time as a 
function of $\tau N^{\alpha}$. The worst error of the data points is 
about the size of the symbol, hence errors were omitted for clarity.
Clearly, the average median time to escape follows the relation:

\beq\label{sim_escape_scaling}
\left< t_{med} \right> = 
	\frac{a N}{\tau} \exp (b \tau N^{\alpha})
\eeq

\noindent
where $~a,b,\alpha~$ are constants (given below).

% ------------------------------------------
% Place Figure 16 around here
% ------------------------------------------

In order to appreciate this result, we need an appropriate variable 
transformation.
Recall that in previous sections we saw that the projection of the noise 
scales as $~ 1/ \sqrt{N}$, hence its second moment scales as $~ 1/N$.
On the other hand, correlations between the noise
terms might affect this scaling, therefore the exponent should be
$~~ b \tau N^{\alpha} \equiv \tau / \epsilon ~$, where $\alpha < 1$.
Our simulations show that $\alpha \approx 0.5$. 
Also note that $b$ increases linearly with $\sigma^{2\alpha}$.
The prefactor is affected by the nature of the sequential scheme,
i.e.\ the fact that time is rescaled.
As expected, it was found that the polynomial increase in the MFPT is 
linear with the size of the system, with $a \approx 0.07$.
The constant slope in the exponent ($b$) found from simulations is 
$\approx 11.2$, while the prediction given by the model is $\approx 9.4$.

\section{Discussion}\label{discussion}

In this work we have studied the time series generated by a noisy 
Sequence Generator (SGen). We have focused on the robustness of the 
isolated analytical results in the presence of noise, 
the issue of phase coherence and escape time from a meta-stable attractor.

Although the system does not becomes ergodic in the presence of noise,
the attractor is broadened. We have analyzed this phenomena for the case
of a perceptron-SGen and found
that the attractor in phase space is inversely broadened as $\sqrt{N}$.
Nevertheless, it is clear that this result is applicable to more 
complicated architectures as well.

Analysis of the phase coherence is highly important in quasi-periodic 
complex time series since, in general, merely identifying the governing 
frequencies in the system is insufficient. 
To investigate this phenomena, we have analyzed the behaviour of 
the diffusion coefficient. It is related to the 
divergence with time of the variance of the phase error. For uncorrelated
noise, we show that the diffusion coefficient should scale inversely with
$N$. In order to test this argument numerically, we used two updating 
schemes. The parallel scheme fits exactly to our model. 
For the sequential scheme the diffusion coefficient scales 
as $1/N^2$ since the time is rescaled by $1/N$. 
Nevertheless, the conclusion is the same, namely, coherence 
is indeed maintained for time length which scales less than linear
with the size of the system, i.e.\, $~t \sim N^a ~~ (a<1) ~$ for large
$N$. The loss of phase coherence is also manifested in a broadening 
of the dominant component in the power spectrum in the same manner,
namely, the larger $N$ is, the sharper the dominant component.

We have calculated numerically the attractor 
dimension from time series that were generated by SGen's
for both cases (noisy and isolated), for the perceptron as well as
for a multi-layer network. The results for the noisy/isolated system
are very similar and in agreement with the analytical results obtained
for the isolated system \cite{sgen1}, i.e.\ the attractor dimension does
not change in the presence of noise. This result is, of course, not 
surprising from the point of view of dynamical systems, as described in
section \ref{chap_AD}.

When the noise interacts with a system that consists of more than a
single attractor, one distinguishes between two time scales.
In the short term, 
the system is still stable with respect to the previous results, 
namely one can work within the framework of a single attractor. 
However for large times, fluctuations take over and the system may 
escape from the initial basin of attraction. 
We have developed the theory for the mean first passage time to escape
an attractor defined by a Fourier component in the power spectrum of the 
weight vector. For this analytical investigation, proper variables were 
identified. These are the amplitudes of the solution to the unperturbed
system. Without noise, we found that these variables are connected via
coupled equations, however, in the generic case only one variable has a
stable non-zero value (above bifurcation). Adding noise to the dynamics
perturbs this solution. We have focused on the case of two symmetric
attractors. In the limit of small noise and not far from the bifurcation
we were able to reduce the dimensionality of the dynamics into a $1D$
flow. This manipulation allows us to use the theory developed for discrete
dynamics driven by noise. The results resemble those obtained in
systems with potential barrier undergoing a tunneling in the sense that
the escape time has a polynomial prefactor and a leading exponential
term. We defined a reduced variable $\tau$ (equation \ref{tau}) which is
closely related to the amplitude of the solution. 
This quantity plays the role of the potential gap. 
Simulations of the SGen with two symmetric attractors have
shown that our theory, and especially the reduction to a $1D$ flow, are
correct. The small corrections to the theory are due to the correlations
between the noise terms in the sequential scheme, while in the theory we
assumed uncorrelated noise.
In order to complete the picture we still have to solve the non-symmetric
case, and to extend it to more than two attractors 
(details will be given in \cite{next_paper}). 
However, we expect that
as long as the number of significant attractors does not scale with the
size of the system, this theory can provide a good explanation.
Further extensions can also be made to the multi-layer network.

Although this analysis was applied to a perceptron-SGen, it is reasonable
to expect that the general properties remain valid in the case of a generic
two-layer network where each perceptron-SGen exhibits its attractors.

%\ack

A.P. would like to thank Y. Ashkenazy for fruitful 
discussions concerning measuring attractor dimension and for providing 
a software. I.K. and D.A.K. acknowledge the partial support of the 
Israel Science Foundation.

%\section*{References}

\end{document}